\documentclass[12pt,preprint]{aastex}

\newcommand{\lea}{{\>\rlap{\raise2pt\hbox{$<$}}\lower3pt\hbox{$\sim$} \>}}
\newcommand{\gea}{{\>\rlap{\raise2pt\hbox{$>$}}\lower3pt\hbox{$\sim$} \>}}

\begin{document}

\title{SIMILARITIES IN POPULATIONS OF STAR CLUSTERS}

\author{S. Michael Fall\/\altaffilmark{1} and 
              Rupali Chandar\/\altaffilmark{2}
              }


\altaffiltext{1}{Space Telescope Science Institute,
         3700 San Martin Drive, Baltimore, MD 21218, USA;
         fall@stsci.edu}
\altaffiltext{2}{Department of Physics and Astronomy,
         University of Toledo, Toledo, OH 43606, USA;
         rupali.chandar@ utoledo.edu}

\begin{abstract}

We compare the observed mass functions and age distributions of star 
clusters in six well-studied galaxies: the Milky Way, Magellanic Clouds, 
M83, M51, and Antennae.
In combination, these distributions span wide ranges of mass and age:
$10^2 \lea M/M_{\odot} \lea 10^6$ and $10^6 \lea \tau/ \mbox{yr} \lea 
10^9$.
We confirm that the distributions are well represented by power laws:
$dN/dM \propto M^{\beta}$ with $\beta \approx -1.9$ and $dN/d \tau 
\propto \tau^{\gamma}$ with $\gamma \approx -0.8$.
The mass and age distributions are approximately independent of each 
other, ruling out simple models of mass-dependent disruption.
As expected, there are minor differences among the exponents, at 
a level close to the true uncertainties,  $\epsilon_{\beta} \sim 
\epsilon_{\gamma} \sim$~0.1--0.2.
However, the overwhelming impression is the similarity of the mass 
functions and age distributions of clusters in these different galaxies, 
including giant and dwarf, quiescent and interacting galaxies.
This is an important empirical result, justifying terms such as 
``universal" or ``quasi-universal."
We provide a partial theoretical explanation for these observations 
in terms of physical processes operating during the formation and 
disruption of the clusters, including star formation and feedback, 
subsequent stellar mass loss, and tidal interactions with passing 
molecular clouds.
A full explanation will require additional information about the 
molecular clumps and star clusters in galaxies beyond the Milky Way. 

\end{abstract}


\section{INTRODUCTION}

Star clusters form in the dense parts (``clumps'') of molecular clouds 
(Lada \& Lada 2003; McKee \& Ostriker 2007). 
They are subsequently destroyed by several processes, beginning 
with the expulsion of residual gas by massive young stars (``feedback''), 
further mass loss from intermediate- and low-mass stars, tidal disturbances 
by passing molecular clouds, and stellar escape driven by internal two-body 
relaxation (Binney \& Tremaine 2008). 
These processes disperse the stars from clusters into the surrounding
stellar field.
Star clusters therefore represent an intermediate stage in the
transformation of the clumpy interstellar medium (ISM) of a galaxy into
a relatively smooth stellar distribution.

In this paper, we show that there are some remarkable similarities in 
the statistical properties of cluster populations in different galaxies.
In particular, we focus on the univariate mass and age distributions,
$\psi(M) \propto dN/dM$ and $\chi(\tau) \propto dN/d\tau$, and the
bivariate mass--age distribution, $g(M,\tau) \propto \partial^2N/
\partial{M}\partial{\tau}$.
These distributions encode valuable information about the formation
and disruption of clusters, for example, whether less massive ones 
dissolve faster than more massive ones.
In the next section, we present the mass and age distributions of the
clusters in six well-studied galaxies, based on published data, but 
now displayed in a uniform manner.
In the following section, we provide a partial theoretical explanation 
for these observations, and in the last section, we summarize and 
place our results in the context of other work in this field. 

As in our previous papers, we use the term ``cluster'' for any 
concentrated aggregate of stars with a density much higher than 
that of the surrounding stellar field, whether or not it also contains 
gas, and whether or not it is gravitationally bound.
This is the standard definition in the star formation community (Lada \&
Lada 2003; McKee \& Ostriker 2007).
Some authors use the term ``cluster'' only for gas-free or gravitationally 
bound objects.  
We reject these definitions for several reasons.
(1)  They ignore the fact that gas-free and gas-rich clusters (including
molecular clumps and HII regions) are the same objects observed in 
different evolutionary phases.
(2) It is impossible to tell from images alone which clusters are 
gravitationally bound (have negative energy) and which are unbound 
(have positive energy).
In principle, spectra would help, but in practice, they are seldom
available and are often contaminated by non-virial motions (stellar 
winds, binary stars, etc).
(3) $N$-body simulations show that unbound clusters retain the 
appearance of bound clusters for many ($\sim$10--50) crossing 
times (Baumgardt \& Kroupa 2007).
By ``disruption,'' we mean the removal of mass from a cluster,
whether this occurs gradually or suddenly, and whether it
leaves the cluster bound or unbound.

\section{MASS AND AGE DISTRIBUTIONS}

Figures~1 and~2 show the mass functions and age distributions of the 
clusters in six nearby galaxies.
The mass functions are plotted for several disjoint age intervals and 
the age distributions for several disjoint mass intervals.
These figures are based on data from Lada \& Lada (2003) for the solar 
neighborhood in the Milky Way, Chandar et~al.\ (2010a) for the Large
and Small Magellanic Clouds (LMC and SMC), Chandar et~al.\ (2010c) 
for a large field in M83, Chandar et~al.\ (2011) for M51, and Fall et~al.\ 
(2009) for the Antennae.
For the clusters in the solar neighborhood, the masses and ages were 
derived from photometry of individual stars; for the clusters in the other 
galaxies, they were derived from integrated photometry in several 
wavebands (usually $UBVI{\rm H}\alpha$) and comparisons with 
stellar population models. 
The H$\alpha$ photometry is crucial for distinguishing clusters
younger and older than $\sim$$10^7$~yr. 
Lada \& Lada (2003) and Chandar et~al.\ (2010b) describe these 
methods in more detail.

For ease of comparison, we have made two simple adjustments 
to the published mass and age distributions when constructing 
Figures~1 and~2.
First, we replotted them in a uniform format: $\log(dN/dM)$ against 
$\log(M/M_{\odot})$ and $\log(dN/d\tau)$ against $\log(\tau/{\rm yr})$.
For the solar neighborhood, the original distributions were presented 
in the form $\log(MdN/d\log{M})$ against $\log(M/M_{\odot})$ and
$\log(dN/d\log{\tau})$ against $\log(\tau/{\rm yr})$.
Second, we adopted a uniform conversion from light to mass based 
on stellar population models with the Chabrier (2003) IMF.
For the LMC, SMC, M51, and Antennae, the original distributions 
were based on models with the Salpeter (1955) IMF, which have
$\Delta\log{M} = 0.2$ and $\Delta\log{\tau} = 0.0$ relative to models
with the Chabrier (2003) IMF.

The observed mass and age distributions are well represented by 
featureless power laws:
\begin{equation}
dN/dM \propto M^{\beta},
\end{equation}
\begin{equation}
dN/d\tau \propto \tau^{\gamma}.
\end{equation}
We list the best-fit exponents and their formal $1\sigma$ errors for 
the 12 mass functions and 10 age distributions in Tables~1 and~2.
The straight lines in Figures~1 and~2 show the corresponding power 
laws.
Both the mean and median exponents for this sample are 
$\beta = -1.9$ and $\gamma = -0.8$, and the standard deviations 
of individual exponents about the means are $\sigma_{\beta} = 
0.15$ and $\sigma_{\gamma} = 0.18$ (with full ranges $-2.24 
\le \beta \le -1.70$ and $-1.05 \le \gamma \le -0.54$).
As a result of stochastic fluctuations in the luminosities and 
colors of clusters, the true uncertainties (errors) in the exponents, 
$\epsilon_{\beta}$ and $\epsilon_{\gamma}$, are usually larger 
than the formal $1\sigma$ errors listed in Tables~1 and~2, with 
typical values $\epsilon_{\beta} \sim \epsilon_{\gamma} \sim$
0.1--0.2 (Fouesneau et~al.\ 2012).\footnote{
In fact, these estimates are lower limits to $\epsilon_{\beta}$ and 
$\epsilon_{\gamma}$ because they neglect likely systematic 
uncertainties and/or variations in the adopted stellar population 
models and extinction curves.
When we make reasonable allowance for these effects, the true
uncertainties increase to $\epsilon_{\beta} \sim \epsilon_{\gamma} 
\sim 0.2$.} 
Since these are similar to the dispersions $\sigma_{\beta}$ and 
$\sigma_{\gamma}$, we cannot tell whether the small differences 
among the exponents are real, although we do expect differences 
at roughly this level, as explained below.  

Figures 1 and 2 show that the mass and age distributions are 
essentially independent of each other.
This follows from the parallelism of the mass functions in different
age intervals and the age distributions in different mass intervals. 
The vertical spacing between the age distributions differs only
because the adopted mass intervals differ, a consequence of the 
different distances, limiting magnitudes, and sample sizes  
among the galaxies.
Thus, we can approximate the bivariate mass--age distribution as
a product of the two univariate distributions:
\begin{equation}
g(M,\tau) \propto \psi(M)\chi(\tau) \propto M^{\beta}\tau^{\gamma}.
\end{equation}
We originally introduced this model for clusters in the Antennae, 
with the speculation that it might also apply to clusters in other 
galaxies (Fall et~al.\ 2005; Fall 2006; Whitmore et~al.\ 2007).
We emphasize that, for each galaxy, our power-law fits are based 
on data in large, but limited, ranges of mass and age, typically 
$\Delta\log{M} \sim \Delta\log{\tau} \sim 2$.
In combination, however, they span even wider ranges, roughly 
$10^2 \la M/M_{\odot} \la 10^6$ and 
$10^6 \la \tau/{\rm yr} \la 10^9$.

The age distribution, in general, reflects the difference between 
the formation and disruption rates of clusters.
We know from observations that the star formation rate has 
remained roughly constant, within factors of 2--3, during the 
past $10^9$~yr in the solar neighborhood (Wyse 2009 and 
references therein), LMC (Harris \& Zaritsky 2009), and SMC 
(Harris \& Zaritsky 2004).
Simulations indicate that this rate has also varied slowly in the 
Antennae for at least the past $10^8$~yr and possibly $10^9$~yr
(Karl et~al.\ 2011).
The formation rates of stars and clusters likely track each other
closely, simply because most stars form in clusters (Lada \& Lada 
2003; McKee \& Ostriker 2007).
A factor of 2--3 variation in the formation rate of clusters between 
$\tau \sim 10^6$~yr and $10^9$~yr would change the exponent 
of the age distribution by only $\Delta\gamma \approx $~0.10--0.17.
This is similar to the dispersion among galaxies and the 
true uncertainty mentioned above ($\sigma_{\gamma} \sim
\epsilon_{\gamma} \sim$ 0.1--0.2) and is much smaller than the
magnitude of the typical exponent, $|\gamma| \approx 0.8$.
Thus, we are confident that most of the observed decline in the 
age distributions is caused by the disruption of clusters rather than
variations in their formation rate.

Two important conclusions now follow directly from Figures~1
and~2 and Equations~(1)--(3).   
First, the shape of the mass function $\psi(M)$ is preserved over
wide ranges of age, although its amplitude, proportional to the 
age distribution $\chi(\tau)$, declines by a large factor.
Thus, there is no evidence for mass-dependent disruption of 
clusters in this sample of well-studied galaxies. 
Second, both the mass and age distributions are remarkably 
similar from one galaxy to another, with differences mostly
attributable to observational uncertainties.
Thus, there is little if any evidence for galaxy-dependent 
disruption of clusters in this sample, which includes giant 
and dwarf, quiescent and interacting galaxies.
This degree of similarity in an astronomical setting probably 
justifies the adjective ``universal.''
However, since we expect and possibly observe minor 
differences among the mass and age distributions in different 
galaxies, we sometimes employ the more modest term
``quasi-universal.''
  
 It is also worth asking what the mass and age distributions 
 would look like if the disruption were mass dependent.
 Figure 3 helps to answer this question. 
 The curves in this diagram are based on a model in which clusters
 form at a constant rate with a power-law initial mass function and 
 are then disrupted gradually at the following mass-dependent rate:
 \begin{equation}
 dM/d\tau = -M/\tau_d(M),
 \end{equation}
 \begin{equation}
 \tau_d(M) = \tau_* (M/10^4 M_{\odot})^k.
  \end{equation}
Boutloukos \& Lamers (2003) claim that this model represents 
an ``empirical disruption law,'' in which different galaxies have
nearly the same exponents, $k \approx 0.6$, but timescales 
$\tau_*$ that differ by more than two orders of magnitude. 
It is therefore both mass dependent and galaxy dependent.
Fall et~al.\ (2009) derived analytical formulae for the mass 
and age distributions for this and other disruption models. 
For the LMC clusters, de~Grijs \& Anders (2006) advocate  
the parameter values $k = 0.6$ and $\tau_* = 8 \times 
10^9$~yr. 
As Figure~3 shows, this model fails by a wide margin to match 
the observed mass and age distributions of the LMC clusters.
Chandar et~al.\ (2010a, 2010c, 2011) present similar comparisons 
and failures for the SMC, M83, and M51 clusters. 

A key point illustrated by Figure~3 is that any mass dependence 
in the disruption rate will introduce features in both the mass and 
age distributions.
These occur at $M_d(\tau) = 10^4 (\tau/\tau_*)^{1/k} M_{\odot}$ 
and  $\tau_d(M) = \tau_* (M/10^4 M_{\odot})^k$, respectively.
For $M \ga M_d(\tau)$ and $\tau \la \tau_d(M)$, the mass function
retains its initial shape and amplitude, while the age distribution
remains flat.
Conversely, for $M \la M_d(\tau)$ and $\tau \ga \tau_d(M)$, the 
mass function becomes shallower, while the age distribution
steepens.
We note that these signatures of mass-dependent disruption are 
even more prominent in the age distribution than in the mass 
function and would be easy to detect if they were present.
As we have already emphasized, there are no such features
in the observed distributions displayed in Figures~1 and~2. 

The mass and age distributions compiled here are based 
on samples of clusters that are complete or unbiased with 
regard to mass and age. 
We caution readers that some of the distributions in the 
literature are based on samples with less suitable selection 
criteria.
The completeness of the samples may be poorly known 
and/or vary with mass and age; clusters may be excluded
if they are located in crowded regions, have uncertain 
photometry, or are somehow deemed to be unbound.
Selection biases such as these can easily cause 
deviations from power laws in the mass and age 
distributions, as demonstrated explicitly by Fall 
et~al.\ (2009) and Chandar et~al.\ (2010a).
Stochastic fluctuations in the luminosities and
colors of clusters are another cause of spurious features 
in these distributions, especially when relatively narrow
mass and age bins are chosen (e.g., the RSG ``gap'' at 
$\tau\approx$~(1--3)~$\times10^7$~yr; Fall et~al.\ 2005; 
Fouesneau et~al.\ 2012).

Some authors also claim to find an exponential 
steepening of the mass function for $M \ga M_c \sim 
10^5 M_{\odot}$ (Larsen 2009).
These claims, however, are based on the absence of 
only a few clusters relative to extrapolated power laws 
and thus have little statistical significance 
(Chandar et~al.\ 2010b).
The mass functions displayed in Figure~1 do not even
hint at such steepening. 
If they did, it would be appropriate to replace the power 
law in Equation~(3) by a Schechter function, giving
$g(M, \tau) \propto M^{\beta}\tau^{\gamma} \exp(-M/M_c)$. 
We note that this mass--age distribution also represents 
mass-independent disruption (for $M_c = {\rm const}$) 
because it is still of separable form, $g(M,\tau) \propto 
\psi(M)\chi(\tau)$.
This illustrates an important point: features in the mass 
function are not by themselves evidence for 
mass-dependent disruption.\footnote{
For $\beta \ge -2$, the mass function requires an upper
cutoff to prevent the total mass of young clusters in a galaxy 
from diverging.
The debate is whether this cutoff occurs at a relatively low 
mass ($M_c \sim 10^5 M_{\odot}$) and whether it has actually 
been detected in the data.
Most of the claimed detections are based on indirect methods, 
and most ignore extinction and uncertainties in the exponents 
$\beta$ and $\gamma$ of the mass and age distributions.}

Bastian et~al.\ (2012) have presented new mass and age 
distributions for clusters in two fields in M83, including 
the one studied by Chandar et~al.\ (2010c).
We have checked that these distributions are mostly 
consistent with those derived by Chandar et~al.\ and 
reproduced here in Figures~1 and~2.
The discrepancies are relatively small and can be attributed 
to selection biases of the type mentioned above.
The Bastian et~al.\ samples are incomplete for young clusters 
of all masses ($\tau \la 10^7$~yr), for older low-mass clusters 
($\tau \ga 10^8$~yr and $M \la 10^4 M_{\odot}$), and for 
clusters in crowded regions.
When we apply these selection biases to the more complete
Chandar et~al.\ sample, we introduce artificial features in the 
mass and age distributions, mimicking those in the Bastian 
et~al.\ distributions. 
We are currently analyzing complete samples of clusters in
these and several more fields in M83 and will present the 
results in a separate paper.

\section{FORMATION AND DISRUPTION PROCESSES}

In this section, we discuss the physical processes most
likely to have shaped the observed mass and age distributions 
shown in Figures~1 and~2.
Before proceeding, we emphasize that our power-law models
for $\psi(M)$, $\chi(\tau)$, and $g(M,\tau)$ are convenient
fitting formulae suggested directly by observations and, as 
such, are valid irrespective of any theoretical interpretation.
It is also worth emphasizing that we currently lack some 
auxiliary information needed to construct a definitive 
explanation for the mass and age distributions.
There are, for example, no surveys of molecular clumps in 
galaxies beyond the Milky Way.
We also lack accurate measurements of the internal structure
of large and complete samples of extragalactic star clusters.
Nevertheless, we have enough theoretical and observational
knowledge to construct a partial explanation of the mass 
and age distributions as follows.

The observed mass functions of molecular clouds and clumps 
in the Milky Way are power laws with exponents $\beta \approx 
-1.7$ (Shirley et~al.\ 2003; Rosolowsky 2005; Mu{\~n}oz et~al.\ 
2007; Wong et~al.\ 2008). 
Because more massive clouds and clumps have longer lifetimes
than less massive ones, they are overrepresented in the samples
from which the mass functions are derived.
Thus, the initial mass function, corrected for this effect, must be 
slightly steeper than the observed one.
The result of this correction is $\beta \approx -2.0$ for clouds and 
clumps, nearly identical to the exponent for clusters (Fall et~al.\ 
2010).
This is what we would expect to find if all the gas in the clumps 
(protoclusters) were converted into stars, i.e., if the efficiency of 
star formation  ${\cal E}$ were 100\%.
In fact, however, the efficiency is much lower: ${\cal E} \approx$ 
20\%--30\% in the dense clumps where clusters form and a factor 
of $\sim$10 lower still in the larger and more diffuse clouds that
contain the clumps (Lada \& Lada 2003).
Thus, to explain why the mass function of clusters is a power
law with exponent $\beta \approx -1.9$, we must explain why 
the efficiency of star formation ${\cal E}$ is independent (or 
nearly so) of the masses of their antecedent  clumps.

Stars form in a protocluster until their output of energy and 
momentum has expelled the remaining gas. 
This stellar feedback potentially includes radiation pressure 
on dust grains, photoionized gas pressure, protostellar 
outflows, main-sequence winds, and supernovae.
It is likely that radiation pressure predominates in massive
protoclusters and that supernovae, because they come after
the other feedback processes, are relatively ineffective at
removing gas (Krumholz \& Matzner 2009; Murray et~al.\ 2010). 
The condition that feedback can accelerate the remaining gas 
to the escape speed then leads to formulae for ${\cal E}(M)$ 
(Fall et~al.\ 2010).
These depend on whether the feedback is closer to the 
energy-driven limit (with negligible radiative losses) or to the
momentum-driven limit (with maximum losses) and the relation
between the escape speed and the masses of the clumps or,
equivalently, their characteristic radius--mass relation, which
we approximate by a power law, $R \propto M^{\alpha}$.
Most of the feedback in the dense protoclusters is momentum 
driven (see Table~1 of Fall et~al.\ 2010), and in this limit, we 
have ${\cal E} \propto  M^{1 - 2\alpha}$.

Fall et~al.\ (2010) showed that there is a strong correlation 
between the radii and masses of molecular clumps, with
$\alpha \approx 0.4$, based on measurements of CS, 
C$^{17}$O, and 1.2 mm dust emission in the independent
surveys by Shirley et~al.\ (2003), Fa{\'u}ndez et~al.\ (2004), 
and Fontani et~al.\ (2005). 
The more recent survey by Wu et~al.\ (2010), based on 
measurements of different molecular tracers, confirms 
this correlation, with $\alpha \approx 0.5$ (see their Figures 
31--34).
These radius--mass relations imply ${\cal E}(M) \approx 
{\rm constant}$ for momentum-driven feedback.
Thus, we have a simple physical explanation for the 
similarity between the mass functions of clumps and 
clusters.
It remains, then, to explain the power-law form of the mass 
function of the clumps in terms of physical processes in the 
turbulent ISM, an important challenge beyond the scope 
of this paper.

Following the expulsion of natal gas from a cluster, it will 
continue to lose mass from its member stars during the 
normal course of stellar evolution (through winds and
other ejecta).
Stellar mass loss itself removes only 10\%--50\% of the 
mass of a cluster (depending on its age, metallicity, and 
stellar IMF), but if the cluster is weakly bound and resides 
in a tidal field, it can be partially or even completely 
disrupted by this process (Chernoff \& Weinberg 1990).
This happens because the cluster cannot maintain a state 
of virial equilibrium (Fukushige \& Heggie 1995).
Clusters with low concentrations, $c  \equiv \log(r_t/r_c) 
\la 0.7$ (where $r_t$ and $r_c$ are the tidal and core radii), 
are particularly vulnerable.
In terms of the dimensionless central potential, this condition
is equivalent to $W_0 \la 0.3$.
We expect such weakly bound clusters to be numerous as
a result of the earlier expulsion of gas by stellar feedback.

The effect of this disruption process on the mass and age
distributions of the clusters depends primarily on their
concentrations---for clusters with the same stellar IMF, 
until two-body evaporation becomes important 
(Chernoff \& Weinberg 1990; Fukushige \& Heggie
1995).   
In particular, the shape of the mass function will be 
preserved, while its amplitude declines, so long as 
$c$ is not correlated with $M$.
This is a reasonable expectation because the star 
formation efficiency ${\cal E}$, which determines the 
relative weakening of the gravitational potential of the 
clusters by stellar feedback, is independent of $M$.
The top panel of Figure~4 shows $c$ plotted 
against $M$ for LMC and SMC clusters from fits of King 
profiles to {\it Hubble Space Telescope} ({\it HST}) 
images by McLaughlin \& van der Marel (2005).
Evidently, there is no correlation between $c$ and $M$, 
consistent with the fact that the mass function has the 
same power-law shape at all ages.
This result is suggestive rather than definitive, however, 
because it involves only a small fraction of clusters in the 
LMC and SMC.\footnote{
The McLaughlin \& van der Marel (2005) sample includes 
most of the LMC and SMC clusters with structural 
parameters derived from {\it HST} images. 
We have compared the values of $r_c$, $r_h$, and $r_t$ 
derived from {\it HST} and ground-based images of the 
same clusters and find large discrepancies, indicating 
that the latter are not accurate enough for our purposes.}

Star clusters can also be disrupted by tidal interactions with
passing molecular clouds.
Spitzer (1958) proposed this mechanism to account for the 
scarcity of clusters older than $\sim$$10^9$~yr in the 
galactic disk.
We follow the comprehensive treatment by Binney \& Tremaine 
(2008) and distinguish two regimes: 
catastrophic, in which clusters are disrupted suddenly by a single 
strong encounter, and diffusive, in which clusters are disrupted 
gradually by a series of weak encounters.
We denote the mass and half-mass radius of the clusters by 
$M$ and $r_h$ and the corresponding quantities for the
perturbers (molecular clouds) by $M_p$ and $r_{hp}$.
The characteristic internal density of the clusters is then given
by $\rho_h = 3M / (8 \pi r_h^3)$.
Furthermore, we denote the mean number of perturbers per 
unit volume by $n_p$ and the RMS dispersion of relative 
velocities between the clusters and perturbers by 
$\sigma_v$.

The timescale for disruption of clusters by tidal interactions
depends on these parameters in the two regimes as follows:
\begin{equation}
\tau_d \propto \frac{\rho_h^{1/2}}{M_p n_p}
\,\,\,\,\,\,\,\,\,\,\,\,\,\,\,~~~~~~~~~~ {\rm (catastrophic~ regime)},
\end{equation}
\begin{equation}
\tau_d \propto \frac{\sigma_v r_{hp}^2 \rho_h}
{M_p^2 n_p}
\,\,\,\,\,\,\,\,\,\,\,\,\,\,\,\,\,~~~~~~~~~~~ {\rm (diffusive~ regime)}.
\end{equation}
In both regimes, $\tau_d$ depends on the properties of the 
clusters only through $\rho_h$.
Thus, this mechanism will not change the shape of the mass
function of the clusters unless there is a correlation between
$\rho_h$ and $M$.

The middle and bottom panels of Figure~4 show the
half-mass radius and density plotted against mass for 
LMC and SMC clusters with {\it HST} images analyzed
by McLaughlin \& van der Marel (2005).
Evidently, there is a positive correlation between 
$r_h$ and $M$ but none between $\rho_h$ 
and $M$, consistent again with the mass function 
preserving its power-law shape.
As before, this result is suggestive rather than definitive, 
because it is based on an incomplete sample of LMC 
and SMC clusters.
A correlation between $\rho_h$ and $M$ would be 
surprising, however, because the tidal field of the host 
galaxy tends to impose the same mean density, 
$\rho_t = 3 M / (4 \pi r_t^3) = 2 (r_h / r_t)^3  \rho_h$, 
on all clusters at the same galactocentric distance, 
independent of $M$.
This constraint permits a correlation between 
$\rho_h$ and $M$ only if there is a compensating 
correlation between $r_h / r_t \approx 0.7 
(r_c / r_t)^{1/2}$ and $M$ and hence between 
$c$ and $M$.
As noted above, such a correlation is neither expected 
nor observed.

The disruption timescale $\tau_d$ in the catastrophic 
regime depends on the properties of the molecular clouds 
(perturbers) only through their mean smoothed-out 
density, $\bar\rho_p = M_p n_p$.
Binney \& Tremaine (2008) estimate that open clusters 
in the solar neighborhood marginally satisfy the criterion
for disruption in the catastrophic regime and have  
$\tau_d \sim 3 \times 10^8$~yr.
This local estimate of $\tau_d$ should also apply to other 
galaxies, after scaling inversely by $\bar\rho_p$, if the 
internal densities $\rho_h$ of the clusters are known or 
assumed to be the same as those in the solar neighborhood.
The situation is more complicated in the diffusive regime
because $\tau_d$ then also depends on $M_p$, $r_{hp}$, 
and $\sigma_v$, quantities that are much harder to 
determine from observations than $\bar\rho_p$. 
In this case, Equation~(7), which was derived for identical 
perturbers, should be revised so that the disruption rate 
$1 / \tau_d$ is a weighted sum over terms with different 
$M_p$ and $r_{hp}$, based on their frequency of 
occurrence in the population of molecular clouds.

Over longer times, low-mass clusters are disrupted by 
stellar escape driven by internal two-body relaxation 
(``evaporation'').
This process, unlike the others discussed here, inevitably
causes a bend in the mass function.  
The mass at which the bend occurs is age dependent,
reaching $M \sim 10^5 M_{\odot}$ at $\tau \sim 10^{10}$~yr,
which provides a natural explanation for the observed
turnover in the mass function of old globular clusters 
(Fall \& Zhang 2001; McLaughlin \& Fall 2008 and 
references therein).
However, we do not expect to observe this feature in the 
mass functions displayed in Figure~1, because in this case, 
the clusters are too massive and too young to have 
experienced much two-body evaporation (as demonstrated 
quantitatively in the papers from which the mass functions 
were taken).

There are some reasons for expecting the mass and age 
distributions of clusters to be similar in different galaxies.
First, the mass functions of molecular clouds are observed
to be broadly similar (Rosolowsky 2005; Blitz et~al.\ 2007; 
Fukui et~al.\ 2008; Wong et~al.\ 2011).\footnote{
Some differences have been reported in the mass functions 
of molecular clouds in different galaxies.
These comparisons are based on data from different
instruments, analyzed in different ways by different authors,
and pertain to different density thresholds and mass ranges.
Thus, the reported differences probably exaggerate the real
differences to some extent.}
If the mass functions of the clumps, which have yet to be
surveyed outside the Milky Way,  are also similar, then 
clusters in different galaxies would form with similar initial 
conditions. 
Second, the disruption by feedback and subsequent stellar 
mass loss, processes internal to the clusters, is expected to 
operate similarly in different galaxies.  
The rate of disruption by passing molecular clouds 
may differ among galaxies, depending on the mean 
smoothed-out density $\bar\rho_p$ and other factors 
(see Equations~(6) and~(7)), but this effect may be too 
weak to be detectable over much of the accessible 
range of ages.
For example, the estimated disruption timescale for 
clusters in the solar neighborhood, $\tau_d \sim 3 \times 
10^8$~yr (Binney \& Tremaine 2008), exceeds the ages 
of all the clusters used to construct the age distribution 
plotted in Figure~2.

Some of the disruption processes discussed here are
likely influenced by the tidal field of the host galaxy.
Since this has a strong inverse dependence on 
distance from the galactic center, one might expect 
the mass and age distributions of the clusters to 
exhibit some spatial variation.
Unfortunately, it is difficult to make reliable
predictions of this effect because it depends on the 
initial concentrations of the clusters, which in turn are
governed by the expulsion of gas by stellar feedback.
In particular, it is not clear whether there would be 
much of a correlation between the initial concentrations 
and the locations of clusters. 
In any case, it is important to note that the characteristic
tidal fields within galaxies---measured, for example, at 
their effective radii---have relatively weak variations
among galaxies (Fall \& Zhang 2001; McLaughlin \& 
Fall 2008).
Thus, although the disruption of clusters may be
influenced by local tidal fields, we do not expect 
this to cause major differences in the mass and age 
distributions when averaged over whole galaxies.
 
\section{DISCUSSION}

We have reexamined the mass and age distributions of
star clusters in the Milky Way, LMC, SMC, M83, M51, and
Antennae galaxies.
These distributions are well represented by power laws: 
$\psi(M) \propto dN/dM \propto M^{\beta}$ with $\beta 
\approx -1.9$ and $\chi(\tau) \propto dN/d\tau \propto 
\tau^{\gamma}$ with $\gamma \approx -0.8$.
Furthermore, the mass and age distributions are 
approximately independent of each other: $g(M,\tau)
\propto \psi(M) \chi(\tau)$.
Since the rates of star and hence cluster formation have 
varied relatively slowly, the more rapid decline of the age 
distributions must be the result of cluster disruption.  
We find no evidence for mass-dependent disruption; indeed,
simple models with $\tau_d \propto M^{0.6}$ fail by wide 
margins to match the observed mass and age distributions.

The mass functions of the clusters are similar to those of the 
molecular clouds and clumps in which they form, requiring 
a nearly constant (but small) efficiency of star formation.
This is an expected consequence of stellar feedback, given
the observed radius--mass relation of the clumps ($R \propto 
M^{\alpha}$ with $\alpha \approx 0.5$).
The clusters are also disrupted by subsequent stellar mass 
loss and passing molecular clouds.
These processes preserve the power-law shape of 
the mass function unless there are correlations between 
the concentration $c$ and $M$ or between the internal 
density $\rho_h$ and $M$ (until two-body evaporation
becomes important).
The available data, while limited, reveal no such 
correlations.

The main message here is the similarity among the mass 
functions and age distributions of clusters in different galaxies, 
strikingly evident in Figures~1 and~2.
The literature on this subject includes some contradictory
claims: for prominent features in the mass and age distributions,
for strong dependencies on mass or differences among galaxies.
We have shown that selection biases are responsible for several 
of these claims and we suspect they are responsible for the 
others.
It is hard to see how selection biases and/or analysis errors would
make the mass and age distributions appear more similar than 
they really are, whereas it is easy to see how such problems could 
introduce spurious differences. 
In any case, since the distributions presented here are based on 
complete or unbiased samples, we are confident that the similarity 
they exhibit is real.

Nevertheless, we do not expect the mass and age distributions 
to be exactly the same in all parts of all galaxies.
As we have emphasized here and in our previous papers, we
expect minor differences in the age distributions as a consequence
of different histories of cluster formation and/or disruption by passing
molecular clouds. 
We would also not be surprised to find minor differences in the 
mass functions of molecular clumps and hence those of young 
clusters.  
Indeed, we find minor differences in the observed distributions; the
best-fit exponents have dispersions $\sigma_{\beta} = 0.15$ and 
$\sigma_{\gamma} = 0.18$.  
The reality of these differences remains questionable, however, 
because $\sigma_{\beta}$ and $\sigma_{\gamma}$ are close to 
the true uncertainties in the exponents, $\epsilon_{\beta} \sim
\epsilon_{\gamma} \sim $~0.1--0.2.

We also expect any differences in the mass and age distributions
to depend on spatial scale.
In small regions ($\la {\rm few} \times10^2$~pc), there may be 
large variations in the formation and disruption rates, and hence 
in the mass and age distributions.
As more of these small regions are combined into larger ones,
the variations will average out, and the differences in the mass
and age distributions will diminish.  
Figures~1 and~2 show the important result that these differences 
are negligible or barely detectable on the scale of whole galaxies.
Evidently, the similarity among the mass and age distributions 
overwhelms such differences. 
This is the sense in which we consider the distributions to be 
``universal'' or ``quasi-universal.''

Finally, we mention several future studies that would help to
advance this subject.
On the observational side, it would be interesting to survey 
the molecular clumps in nearby galaxies, to derive their mass 
functions, radius--mass relations, and other properties, for 
comparisons with those in the Milky Way.
Another important goal is to measure the internal structure of 
clusters from {\it HST} images (especially $c$ and $\rho_h$) 
in large unbiased samples in several nearby galaxies.
From our experience, H$\alpha$ measurements are necessary
for accurate determinations of the mass and age distributions 
of young clusters ($\tau \la 3 \times 10^7$~yr) and should be 
included in all future studies.
On the theoretical side, it would be interesting to explore 
further the effects of stellar mass loss on the evolution, 
stability, and disruption of weakly-bound clusters in tidal 
fields.

\acknowledgements
We thank Mark Krumholz, Chris Matzner, and Brad Whitmore 
for helpful comments.
S.M.F. acknowledges support from NASA grant AR-09539.1-A,
awarded by the Space Telescope Science Institute, which is 
operated by AURA, Inc., under NASA contract NAS5-26555.  
R.C. acknowledges support from the NSF through CAREER
award 0847467.

\clearpage

\begin{figure}
\epsscale{0.8}
\plotone{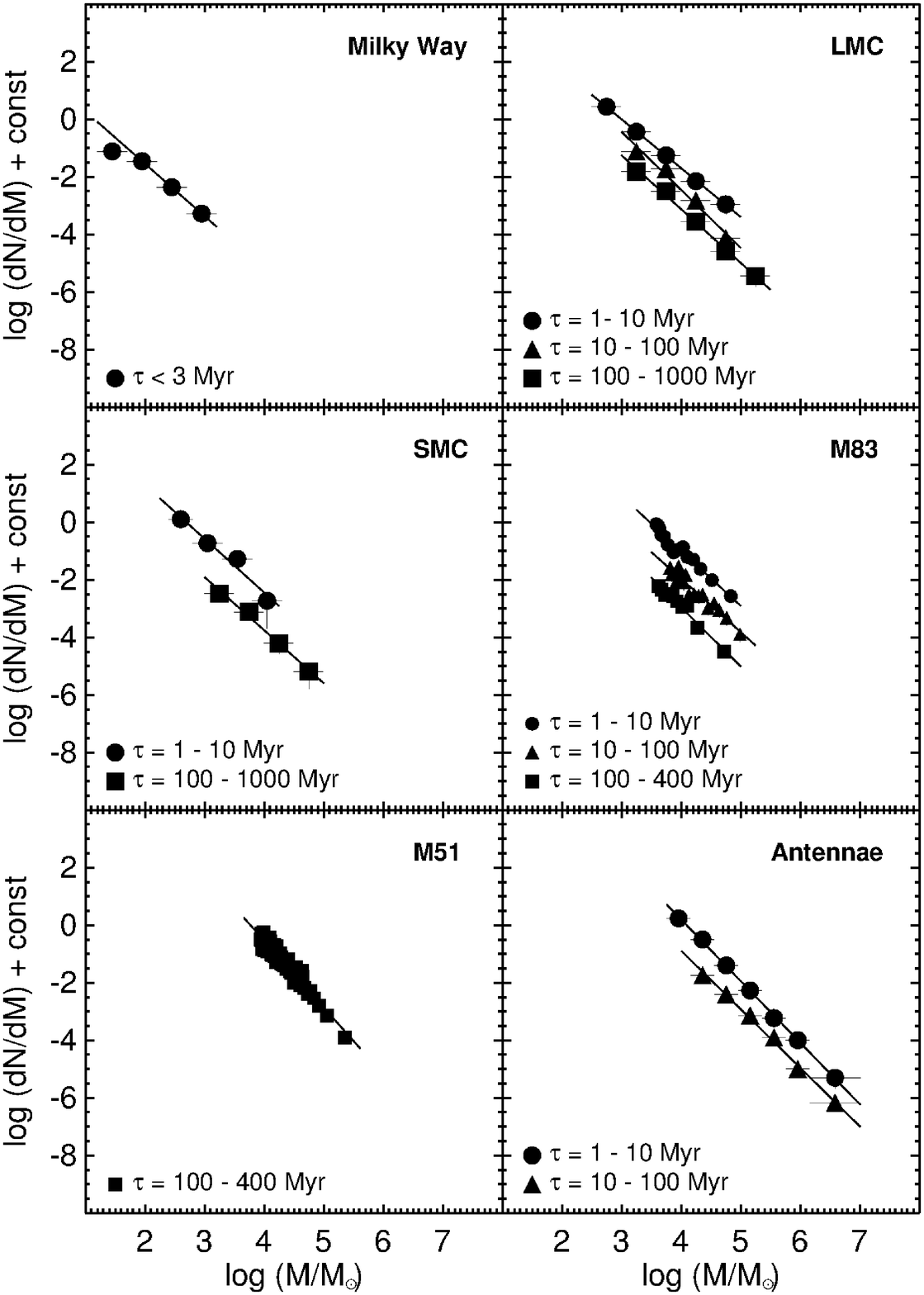}
\caption{Mass functions of star clusters in different age intervals in 
different galaxies (as indicated).  
These have been adapted from the references given in the text.
The absolute normalizations of the mass functions are arbitrary, 
but the relative normalizations within each panel are preserved.
The lines show power laws, $dN/dM \propto M^{\beta}$, with the 
best-fit exponents listed in Table~1.
Note that these are all close to $\beta = -1.9$.
}
\end{figure}

\begin{figure}
\epsscale{0.8}
\plotone{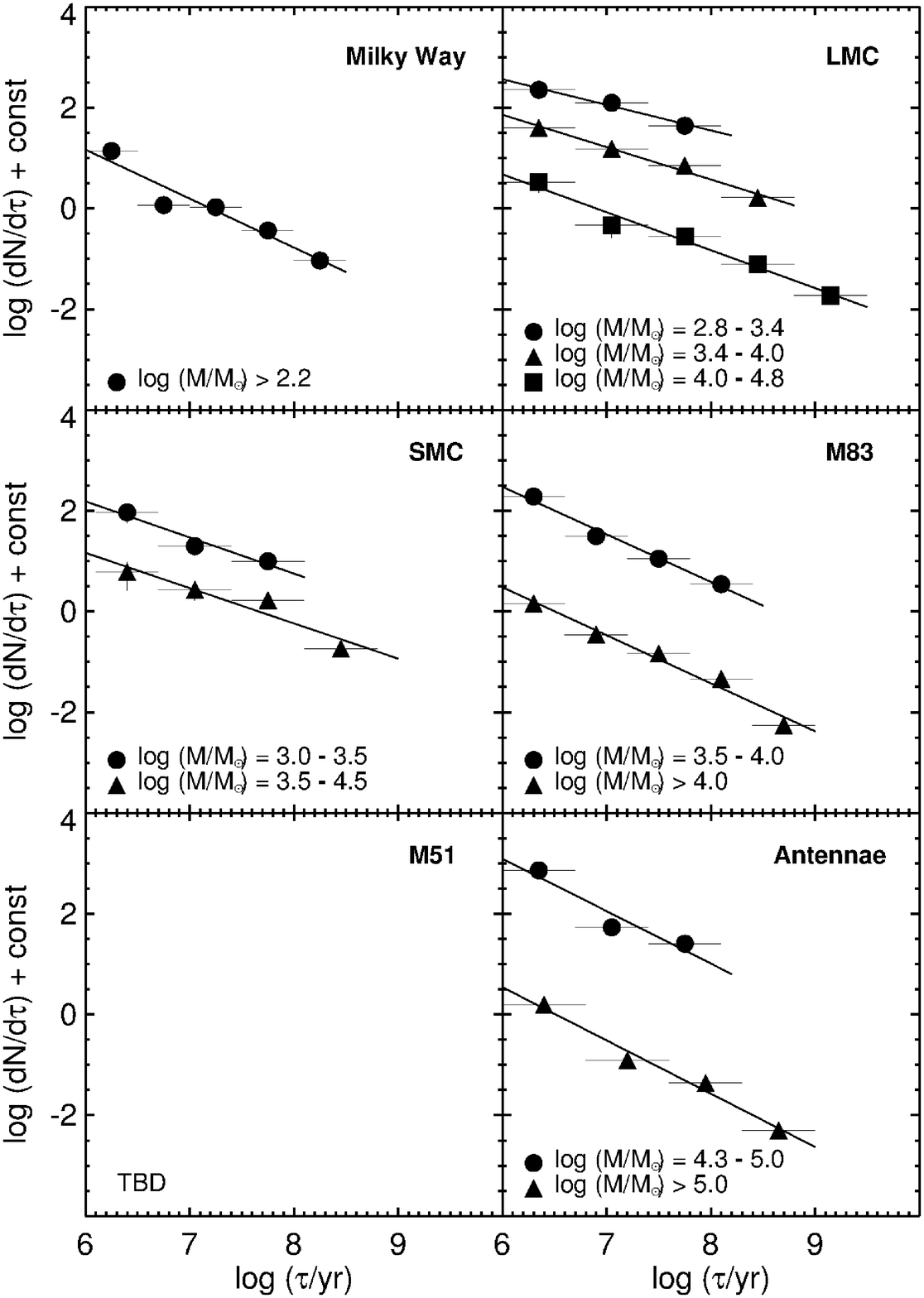}
\caption{Age distributions of star clusters in different mass intervals
in different galaxies (as indicated).
These have been adapted from the references given in the text.
The absolute normalizations of the age distributions are arbitrary, 
but the relative normalizations within each panel are preserved.
The vertical spacing between the age distributions depends 
on the adopted mass intervals, which differ among the galaxies
for practical reasons (distance, limiting magnitude, sample size).
The lines show power laws, $dN/d\tau \propto \tau^{\gamma}$, 
with the best-fit exponents listed in Table~2.
Note that these are all close to $\gamma = -0.8$.
}
\end{figure}

\begin{figure}
\epsscale{1.0}
\plotone{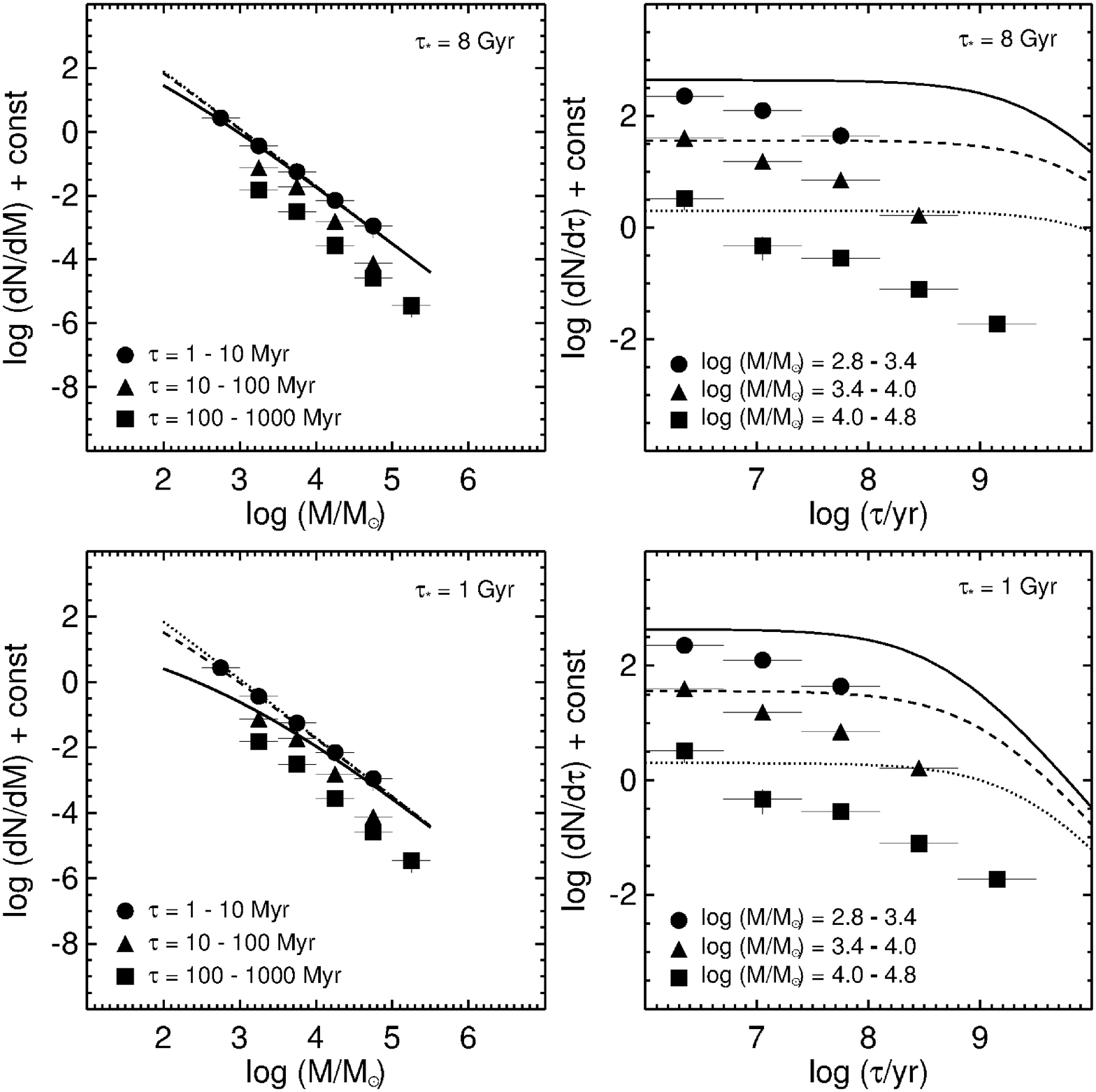}
\caption{
Mass functions and age distributions of star clusters in the LMC (left-hand 
and right-hand panels, respectively).
The data points are the same as in Figures~1 and 2.
The curves are based on a simple model with gradual mass-dependent 
disruption, computed from Equation~(12) of Fall et al. (2009) with 
$\beta = -2$, $k=0.6$, and either $\tau_*=8$~Gyr (top panels) or 
$\tau_*=1$~Gyr (bottom panels).
The three curves in each panel correspond to the same age and mass 
intervals as the data points.
Note that the models are poor fits to all the data, except the mass 
function at the youngest ages ($\tau \le 10^7$~yr).
}
\end{figure}

\begin{figure}
\epsscale{0.55}
\plotone{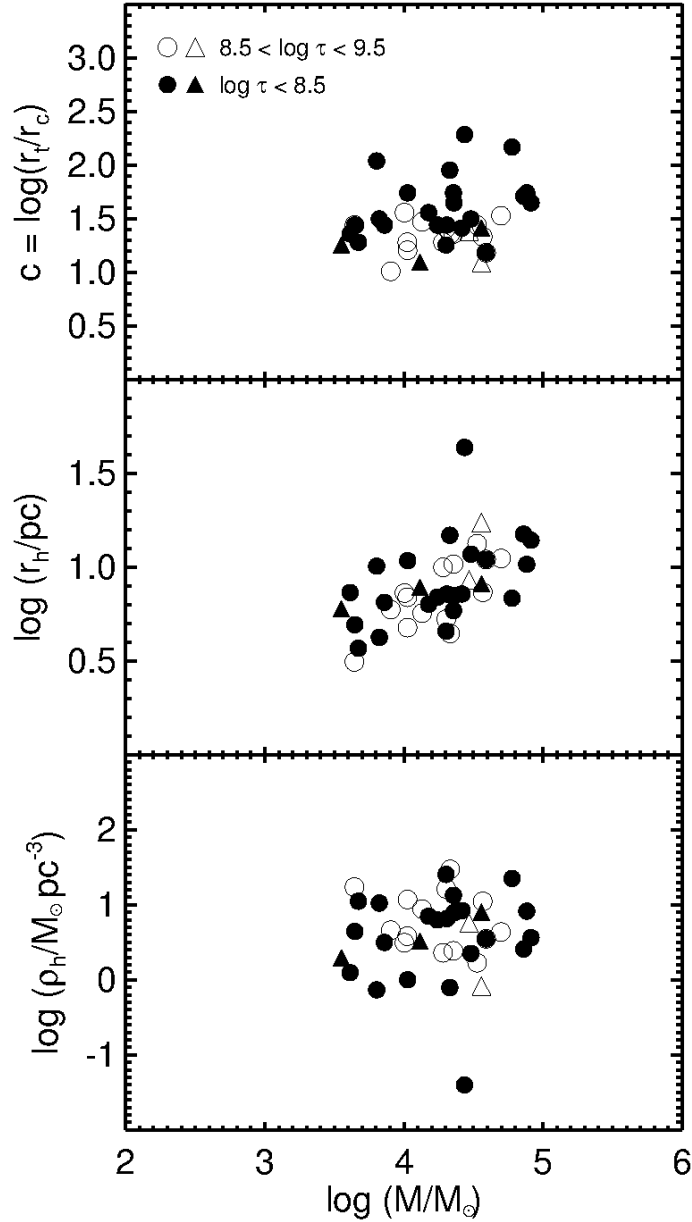}
\caption{
Concentration, half-mass radius, and half-mass density plotted against
mass for LMC clusters (circles) and SMC clusters (triangles) in
two intervals of age (as indicated).
All quantities were determined by McLaughlin \& van der Marel (2005):
$c$, $r_h$, and $\rho_h$ from fits of King models to {\it HST} images
and $M$ from total luminosities and mass-to-light ratios inferred from 
integrated colors and stellar population models.
Note that there is a positive correlation between $r_h$ and $M$
but none between $c$ and $M$ or between $\rho_h$ and $M$.
}
\end{figure}


\clearpage
\begin{deluxetable}{cccccccl}
\tablecolumns{3}
\tablecaption{Exponents of Mass Functions\label{betavalues}}
\tablewidth{0pt}
\tablehead{
\colhead{Galaxy}  & \colhead{Age Interval} & \colhead{$\beta$\tablenotemark{1}} \\
\colhead{} & \colhead{log($\tau$/yr)} & \colhead{}
}
\startdata
Milky Way  & \enspace$<$6\rlap{.5} & $-1.82 \pm 0.01$     \\
LMC &  6$-$7 & $-1.70\pm0.02$      \\
LMC & 7$-$8 & $-2.02\pm0.23$      \\
LMC & 8$-$9 & $-1.87\pm0.07$      \\
SMC &  6$-$7 & $-1.87\pm0.25$   \\
SMC & 8$-$9  & $-1.84\pm0.14$   \\
M83 & 6$-$7 & $-1.91\pm0.10$   \\
M83 &  7$-$8 & $-1.86\pm0.13$   \\
M83 &  8$-$8\rlap{.6} & $-2.06\pm0.13$   \\
M51 &  8$-$8\rlap{.6} & $-2.24\pm0.08$   \\
Antennae & 6$-$7  & $-2.14\pm0.03$   \\
Antennae & 7$-$8  & $-2.03 \pm0.07$    \\ \hline
\enddata
\tablenotetext{1}{{}Least-squares fits to $\log(dN/dM) = \beta\log{M} + {\rm const}.$}
\end{deluxetable}

\clearpage
\begin{deluxetable}{cccccccl}
\tablecolumns{3}
\tablecaption{Exponents of Age Distributions\label{gammavalues}}
\tablewidth{0pt}
\tablehead{
\colhead{Galaxy} & \colhead{Mass Interval}  & \colhead{$\gamma$\tablenotemark{1}} \\
\colhead{} & \colhead{log($M/M_{\odot}$)}  & \colhead{}
}
\startdata
Milky Way & $>2.2$ & $-0.97\pm0.16$     \\
LMC & 2.8$-$3.4 & $-0.54\pm0.10$    \\
LMC & 3.4$-$4.0  & $-0.68\pm0.07$     \\
LMC & 4.0$-$4.8 & $-0.75\pm0.07$      \\
SMC & 3.0$-$3.5 & $-0.73\pm0.17$    \\
SMC & 3.5$-$4.5 & $-0.70\pm0.16$     \\
M83 & 3.5$-$4.0 & $-0.94\pm0.09$    \\
M83 & $>4.0$ & $-0.95\pm0.09$    \\
Antennae & 4.3$-$5.0  & $-1.04\pm0.30$  \\ 
Antennae & $>5.0$  & $-1.05\pm0.11$   \\ \hline
\enddata
\tablenotetext{1}{{}Least-squares fits to $\log(dN/d\tau) = \gamma\log\tau\break + {\rm const}.$}
\end{deluxetable}


\newpage
\begin{thebibliography}{}

\bibitem{}
Bastian, N., Adamo, A., Gieles, M., et al. 2012, MNRAS, 419, 2606

\bibitem{}
Baumgardt, H., \& Kroupa, P. 2007, MNRAS, 380, 1589

\bibitem{}
Binney, J., \& Tremaine, S. 2008, Galactic Dynamics (2nd ed.; Princeton, NJ: 
Princeton Univ. Press)

\bibitem{}
Blitz, L., Fukui, Y., Kawamura, A., et al.
2007, in Protostars and Planets V, ed. B. Reipurth, D. Jewitt, \& K. Keil 
(Tucson, AZ: Univ. Arizona Press), 81

\bibitem{}
Boutloukos, S. G., \& Lamers, H. J. G. L. M. 2003, MNRAS, 338, 717

\bibitem{}
Chabrier, G. 2003, PASP, 115, 763

\bibitem{}
Chandar, R., Fall, S. M., \& Whitmore, B. C. 2010a, ApJ, 711, 1263

\bibitem{}
Chandar, R., Whitmore, B. C., Calzetti, D., et al. 2011, ApJ, 727, 88

\bibitem{}
Chandar, R., Whitmore, B. C., \& Fall, S. M. 2010b, ApJ, 713, 1343

\bibitem{}
Chandar, R., Whitmore, B. C., Kim, H., et al. 2010c, ApJ, 719, 966

\bibitem{}
Chernoff, D. F., \& Weinberg, M. D. 1990, ApJ, 351, 121

\bibitem{}
de Grijs, R., \& Anders, P. 2006, MNRAS, 366, 295

\bibitem{}
Fall, S. M. 2006, ApJ, 652, 1129

\bibitem{}
Fall, S. M., Chandar, R., \& Whitmore, B. C. 2005, ApJ, 631, L133 

\bibitem{}
Fall, S. M., Chandar, R., \& Whitmore, B. C., 2009 ApJ, 704, 453 

\bibitem{}
Fall, S. M., Krumholz, M. R., \& Matzner, C. D. 2010, ApJ, 710, L142

\bibitem{}
Fall, S. M., \& Zhang, Q. 2001, ApJ, 561, 751

\bibitem{}
Fa{\'u}ndez, S., Bronfman, L., Garay, G., et al. 2004, A\&A, 426, 97

\bibitem{}
Fontani, F., Beltr{\'a}n, M. T., Brand, J., et al. 2005, A\&A, 432, 921 

\bibitem{}
Fouesneau, M., Lancon, A., Chandar, R., \& Whitmore, B. C. 2012, 
ApJ, 750, 60

\bibitem{}
Fukui, Y., Kawamura, A., Minamidani, T., et al. 2008, ApJS, 178, 56

\bibitem{}
Fukushige, T., \& Heggie, D. C. 1995, MNRAS, 276, 206

\bibitem{}
Harris, J., \& Zaritsky, D. 2004, AJ, 127, 1531

\bibitem{}
Harris, J., \& Zaritsky, D. 2009, AJ, 138, 1243

\bibitem{}
Karl, S. J., Fall, S. M., \& Naab, T. 2011, ApJ, 734, 11

\bibitem{}
Krumholz, M. R., \& Matzner, C. D. 2009, ApJ, 703, 1352

\bibitem{}
Lada, C. J., \& Lada, E. A. 2003, ARA\&A, 41, 57

\bibitem{}
Larsen, S. S. 2009, A\&A, 494, 539

\bibitem{}
McKee, C. F., \& Ostriker, E. C. 2007, ARA\&A, 45, 565

\bibitem{}
McLaughlin, D. E., \& Fall, S. M. 2008, ApJ, 679, 1272

\bibitem{}
McLaughlin, D. E., \& van der Marel, R. P. 2005, ApJS, 161, 304

\bibitem{}
Mu{\~n}oz, D. J., Mardones, D., Garay, G., et al. 2007, ApJ, 
668, 906

\bibitem{}
Murray, N., Quataert, E., \& Thompson, T. A. 2010, ApJ, 709, 191

\bibitem{}
Rosolowsky, E. 2005, PASP, 117, 1403

\bibitem{}
Salpeter, E. E. 1955, ApJ, 121, 161

\bibitem{}
Shirley, Y. L., Evans, N. J., Young, K. E., Knez, C., \& Jaffe, D. T.
2003, ApJS, 149, 375

\bibitem{}
Spitzer, L. 1958, ApJ, 127, 17

\bibitem{}
Whitmore, B. C., Chandar, R., \& Fall, S. M. 2007, AJ, 133, 1067 

\bibitem{}
Wong, T., Hughes, A., Ott, J., et al. 2011, ApJS, 197, 16

\bibitem{}
Wong, T., Ladd, E. F., Brisbin, D., et al. 2008, MNRAS, 386, 1069

\bibitem{}
Wu, J., Evans, N. J., Shirley, Y. L., \& Knez, C. 2010, ApJS, 188, 313

\bibitem{}
Wyse, R. F. G. 2009, in IAU Symp. 258, The Ages of Stars, ed. 
E. E. Mamajek, D. R. Soderblom, \& R. F. G. Wyse (Cambridge: 
Cambridge Univ. Press), 11

\end{thebibliography}
\end{document}